\documentclass[manuscript,screen, nonacm]{acmart}
\usepackage{algorithm}
\usepackage{algpseudocode}
\usepackage{array}
\usepackage{subcaption}
\usepackage{enumitem}
\usepackage{multirow}

\renewcommand\footnotetextcopyrightpermission[1]{} 
\setcopyright{none}
\AtBeginDocument{%
  }

\begin{document}

\title[Reflection Before Action]{Reflection Before Action: Designing a Framework for Quantifying Thought Patterns for Increased Self-awareness in Personal Decision Making}

\author{Morita Tarvirdians}
\email{M.Tarvirdians@tudelft.nl}
\orcid{0000-0003-4246-0016}
\email{}
\affiliation{%
  \institution{TU Delft}
  \city{Delft}
  \state{}
  \country{The Netherlands}
}
\author{Senthil Chandrasegaran}
\email{r.s.k.chandrasegaran@tudelft.nl}
\orcid{0000-0003-0561-2148}
\email{}
\affiliation{
   \institution{TU Delft}
  \city{Delft}
  \state{}
  \country{The Netherlands}
}
\author{Hayley Hung}
\email{h.hung@tudelft.nl}
\orcid{0000-0001-9574-5395}
\email{}
\affiliation{%
  \institution{TU Delft}
  \city{Delft}
  \state{}
  \country{The Netherlands}
}
\author{Catholijn M. Jonker}
\email{c.m.jonker@tudelft.nl}
\orcid{0000-0003-4780-7461}
\email{}
\affiliation{%
  \institution{TU Delft/Leiden University}
  \city{Delft/Leiden}
  \state{}
  \country{The Netherlands}
}
\author{Catharine Oertel}
\email{c.r.m.m.oertel@tudelft.nl}
\orcid{0000-0002-8273-0132}
\email{}
\affiliation{%
  \institution{TU Delft}
  \city{Delft}
  \state{}
  \country{The Netherlands}
}

\renewcommand{\shortauthors}{Tarvirdians et al.}

\begin{abstract}
When making significant life decisions, people increasingly turn to conversational AI tools, such as large language models (LLMs). However, LLMs often steer users toward solutions, limiting metacognitive awareness of their own decision-making. In this paper, we shift the focus in decision support from solution-orientation to reflective activity, coining the term pre-decision reflection (PDR). We introduce PROBE, the first framework that assesses pre-decision reflections along two dimensions: breadth (diversity of thought categories) and depth (elaborateness of reasoning).
Coder agreement demonstrates PROBE’s reliability in capturing how people engage in pre-decision reflection. Our study reveals substantial heterogeneity across participants and shows that people perceived their unassisted reflections as deeper and broader than PROBE’s measures. By surfacing hidden thought patterns, PROBE opens opportunities for technologies that foster self-awareness and strengthen people’s agency in choosing which thought patterns to rely on in decision-making.
\end{abstract}





\begin{teaserfigure}
  \includegraphics[width=\textwidth]{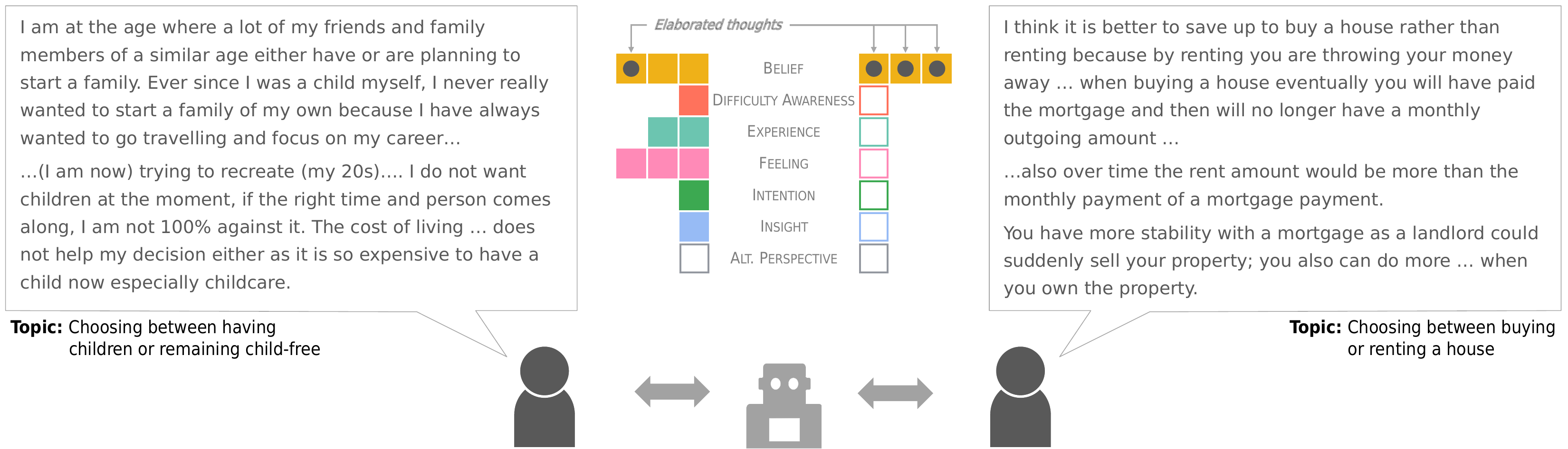}
  \label{fig:teaser}
  \vspace{-5mm}
  \caption{Examples of two pre-decision reflections generated by participants (left and right) through interaction with a conversational agent (center). PROBE enables identification and evaluation of thought patterns each participant demonstrates in their reflection, based on seven pre-defined aspects. Through PROBE, the difference in patterns between the two reflections are shown. The presence of different aspects of reflection is shown via filled squares, while elaboration of thought under each aspect is shown via filled circles. The person on the left shows superficial consideration of various aspects, whereas the person on the right shows deep fixation on one aspect.}
  \Description{The figure is split into two distinct sections, one on the left and one on the right, each representing a separate pre-decision reflections generated by participants. The central column between the two sections lists 7 PROBE categories: Belief, Awareness of Difficulties, Experience, Feeling, Intention, Insight and Alternative Perspective. 
  The text of left reflection: I am at the age where a lot of my friends and family members of a similar age either have or are planning to start a family.
    Ever since I was a child myself, I never really wanted to start a family of my own because I have always wanted to go travelling and focus on my career...
    ...(I am now) trying to recreate (my 20s)... I do not want children at the moment, if the right time and person comes along, I am not 100
    The cost of living does not help my decision either as it is so expensive to have a child now especially childcare.
    The text of right reflection: I think it is better to save up to buy a house rather than renting because by renting you are throwing your money away ... when buying a house eventually you will have paid the mortgage and then will no longer have a monthly outgoing amount
    ...also over time the rent amount would be more than the monthly payment of a mortgage payment.
    You have more stability with a mortgage as a landlord could suddenly sell your property; you also can do more when you own the property.
    The central columns, illustrating PROBE categories and the elaborated thoughts shows that the reflection on the left side includes various thought categories: 3 Belief (only 1 elaborated), 1 Awareness of difficulties, 2 Experience, 3 Feeling, 1 Intention, 1 Insight
    On the other hand, the right reflection only has 3 Beliefs but all elaborated.}
\end{teaserfigure}

\maketitle

\section{Introduction}
Decision making is a constant in human life, with adults making an estimated 35,000 choices each day~\cite{pignatiello2020decision}.
While many are trivial, others, such as whether to quit a job or have children, carry lasting consequences and require careful consideration.
To support their decision making, people often turn to friends or decision support systems and, since the release of ChatGPT \cite{openai2023gpt4}, increasingly to large language models (LLMs) \cite{altman2025}.
Such systems, however, focus on outcomes rather than the process of decision making. 
The optimization of LLMs through RLHF \cite{ouyang2022training} rewards helpfulness and user satisfaction, resulting in a tendency to produce solution-centric responses.
Yet such answers are based on a shallow understanding of the situation and risk undermining the user’s agency in the decision process, in some cases leading to irreversible and harmful outcomes \cite{guardian2025suicide}.

One of the challenges of decision making is keeping oneself aware of unconscious biases and erroneous thinking.
This awareness is brought about through reflection, which is defined as \textit{``Active, persistent and careful consideration of any belief or supposed form of knowledge in the light of the grounds that support it, and the further conclusions to which it tends''}~\cite[p.\ 6]{dewey1910how}.
While there are systems that support reflection, they tend to 
focus retrospectively on the process of decision making.
They help users make sense of past decisions and the factors that shaped them. 
Such systems are used most often in behavior change support, for example, in reflecting on daily dietary choices to encourage healthier eating~\cite{foodJournal, foodJournal2, foodsJournal3}.

However, making people aware of their own biases in thought patterns is critical prior to making significant decisions, as people are subject to well-documented cognitive limitations in decision making~\cite{sweller1988cognitive, tversky1974judgment, kruger1999unskilled, korte_biases}. One such limitation is bounded awareness~\cite{bounded1,bounded}, where individuals focus narrowly on a single category of thought, without realizing the restrictiveness of their perspective.
To our knowledge, no method exists to make people aware of---or measure the quality of---the thought patterns that shape their decisions \textit{before} they are made. This gap limits the design of systems that could support richer reflection before decisions are made. 

We introduce the term \textsc{Pre-Decision Reflection} (PDR) to address the above challenges and knowledge gaps, and develop a framework for assessing PDR,
laying the foundation for future systems that can mirror thought patterns and foster greater awareness and agency in decision making.
We call this method PROBE: a multi-dimensional approach to capture both the diversity of thoughts and the degree to which a person explores a thought category in a pre-decision reflection.
We draw from previously established models for reflection assessment~\cite{wong, ullmann, boud} to create PROBE, and iteratively refine it through a formative study with six participants whose pre-decision reflections are assessed by three coders.

We evaluate the final version of PROBE using pre-decision reflections from participants ($n=40$) collected through their conversations with a conversational agent, related to particular decisions. 
The results demonstrated the reliability of PROBE and its ability to capture participants’ unique thought profiles. 
Analysis revealed substantial variability in thought patterns, in terms of both diversity and depth, and indicated that most participants could benefit from tailored support in enriching these dimensions.
The study also revealed differences between participants' own perceptions of the depth and breadth of their reflections and the depth and breadth measured with PROBE.
While the majority of participants welcomed receiving support during the reflection process, their responses to a post-reflection questionnaire showed a lack of explicit acknowledgment of potential narrowness or limitations in their reflections.
We conclude by discussing potential implications 
for the design of future systems aimed at supporting and augmenting human decision-making in complex contexts.

\section{Related Work}
In this section, we review how reflection has been conceptualized and supported in previous research, as well as how the reflective process has been assessed in prior studies.
\subsection{Reflection}
Schön’s framing of reflection \cite{schon} has been widely adopted in HCI research \cite{review1}. He distinguishes between two types of reflection: reflection-in-action and reflection-on-action. Reflection-in-action refers to the process of reflecting during task execution, allowing individuals to evaluate the situation and adjust decisions in real time \cite{reflection-in-action}. Reflection-on-action, in contrast, is a post hoc process that enables individuals to reconstruct and analyze past events.

A significant body of work in the HCI community has focused on supporting such reflection through technology, with the goal of enhancing self-knowledge and awareness by encouraging reflection on past actions. For example, personal informatics systems provide individuals with insights into their past behaviors by allowing them to review personal data \cite{PIS}, such as sleep patterns \cite{sleeptight}, thereby facilitating reflection-on-action.

Sometimes, the purpose of retrospective reflection extends beyond making sense of the past to informing future choices and actions. For example, in behavior change support systems, individuals iteratively review and reflect on their past actions and results to guide future plans and decisions, as seen, for example, in systems designed to support physical activity \cite{activity, activity2, activity3}.
This type of reflection is sometimes referred to as reflection-for-action \cite{killion}, to show forward-looking orientation of the outcome of the reflection.

Such reflection is particularly useful for decisions made on a regular or daily basis, as it informs future choices by drawing on the awareness gained about past decision making process and its outcome.
On the other hand, non-recurring and high-stakes personal decisions, such as deciding to get married, often lack prior experiences to build upon, and even when past experiences exist, new circumstances typically demand fresh contemplation before committing to a choice.
This type of contemplation, unlike reflection-for-action, is not grounded in a specific past experience but instead draws on the individual’s personal context and circumstances, yet it similarly has a forward-looking orientation. 
As its focus lies in making sense of personal circumstances and thoughts rather than rushing to a decision, we term this process \textit{Pre-Decision Reflection (PDR)}.

\subsection{Reflection Assessment}
Various approaches have been employed to assess reflective practices in the HCI community. 
Some studies have focused on evaluating the outcomes of the reflection. For example, researchers have examined its influence on coaching \cite{inward}, well-being insights \cite{pearl}, perceived self-efficacy \cite{selvreflect}, and perceived well-being \cite{emotionaldepth}. 
Other studies, rather than emphasizing reflection outcomes, have evaluated the effectiveness of the systems that facilitate reflection. This has been done using generic measures such as the System Usability Scale (SUS) \cite{SUS} in \citet{Grayscale} and \citet{participatory}, or with specialized instruments designed to assess reflection support, such as the Technology-Supported Reflection Inventory (TSRI) \cite{TSRI} in \citet{selvreflect} and the Reflection, Rumination, and Thought in Technology (RRTT) scale \cite{RRTscale} in \citet{yin2025travelgalleria}.
However, as noted in surveys by Baumer et al. \cite{review1} and Bentvelzen et al. \cite{revisting}, there remains a lack of research directly analyzing the reflections themselves. Consequently, no established criteria exist for assessing the reflective thoughts or the effectiveness of support systems in enriching them.

\label{sec:evalFrameworks}
The assessment of reflections themselves has been explored in a distinct body of research, including the examples in Table \ref{table:models} and the studies of \cite{kember2008, emotionaldepth, RM1, RM2, RM3, RM4, boud}. These studies primarily aim to identify an individual’s current level of reflection and to explore how individuals can become more reflective. To this end, reflections are typically assessed along two key dimensions: breadth and depth.

The quality of breadth \cite{moon} refers to the extent to which reflections draw on a diversity of angles to consider the subject of reflection.
For example, Ullmann et al. \cite{ullmann} identify seven breadth categories (see Table \ref{table:models}).
The category "Feelings", for instance, indicates that the person reflects on the feelings related to the subject.
Whereas "Perspective", for example, indicates that the reflector examines the subject from a viewpoint different from their own.

The quality of depth \cite{moon}, on the other hand, is one label used to describe the overall depth of a reflection.
Reflections that only describe personal thoughts are considered more superficial than those that include further analysis or interpretation of those thoughts.
For instance, Wong et al. \cite{wong} categorize individuals as critical reflectors, the deepest level in their model, when in addition to the description of their thoughts, they are able to achieve a new interpretation of them by examining the reflection subject from a completely new perspective beyond their own.
\begin{table}[h!]
\newcolumntype{L}[1]{>{\raggedright\let\newline\\\arraybackslash\hspace{0pt}}p{#1}}
\centering
\begin{tabular}{L{2.5cm}L{12cm}} 
\toprule
\textbf{Author}  & \textbf{Model} \\ 
\midrule

\citet{mezirow1990} & \textit{Breadth:} habitual action, thoughtful action, introspection, reflection about content, reflection about process, and reflection about the premises

\textit{Depth:} Non-Reflective, Reflective 
\vspace{0.5em}\\ 
\cmidrule(lr){1-2}
\citet{wong} & \textit{Breadth:} Attending to feelings, Association, Integration, Validation, Appropriation, and Outcome of reflection

\textit{Depth:} Non-reflector, reflector, critical reflector
\vspace{0.5em}\\
\cmidrule(lr){1-2}
\citet{poldner} & \textit{Breadth:} Description, Evaluation, Justification, Dialogue, and Transfer \vspace{0.5em}\\
\cmidrule(lr){1-2}
\citet{prilla} & \textit{Breadth:} Description of an experience, Linking an experience explicitly to other experiences, Linking an experience to knowledge, Responding to the explanation of an experience by providing alternative perspectives, Responding to the explanation of an experience by challenging or supporting assumptions, Contributing to work on a solution by providing reason for the issue, Contributing to work on a solution by providing solution proposals, Showing insights or learning from reflection by describing better individual understanding, Showing insights or learning from reflection by generalising from reflection, Describing or implementing change.

\textit{Depth:} Provision and description of experience, Reflection on experience, Learning or change\\
\cmidrule(lr){1-2}
\citet{ullmann} & \textit{Breadth:} Description of an experience, Feelings, Personal, Critical stance, Perspective, and Outcome (Retrospective and Prospective) 

\textit{Depth:} Non-reflective, Reflective \vspace{0.5em}\\
\bottomrule
\end{tabular}
\caption{Reflection Evaluation Models. The models highlight a shared structure for assessing reflections along breadth and depth, with variations in terminology and categories reflecting the context or purpose of each model—for example, Prilla and Renner focuses on work-based reflection, whereas Ullmann targets educational reflection. Despite these differences, the models converge in treating depth as a hierarchical construct (e.g., non-reflective → reflective → critical reflection), which serves as the primary indicator of reflection quality, while breadth categories function as the building blocks that support and enable such depth.}
\label{table:models}
\end{table}

In studies that employ these models~\cite[e.g.,][]{ip, emotionaldepth, zhang2024}, reflection is typically assessed as a whole, rather than by examining individual thoughts. Depth is often regarded as the ultimate goal that a reflector should achieve in order to demonstrate "good" reflection, while breadth categories are treated as components that collectively contribute to such depth. This approach is valuable in educational settings, where assessment can both evaluate reflective ability and encourage students to become more reflective.

However, this type of assessment is less suitable for pre-decision reflections. In this context, the goal of assessment should not be to assign a quality score (e.g., "You are (not) reflective") but rather to enhance an individual’s understanding of their own thought patterns. Recognizing one’s patterns of thoughts during decision making can foster metacognitive knowledge \cite{flavell1979, schraw1998, lai2011}, as it involves understanding how one thinks and reasons. Active use of this awareness can promote metacognitive regulation \cite{schraw1998, lai2011}, facilitating deeper pre-decision reflection and more deliberate decisions. Moreover, access to individuals’ reflective thought patterns can enable support systems to provide tailored interventions to enhance metacognition (thinking about thinking) \cite{flavell1979}, and pre-decision reflection.

To address this gap, the present paper aimed to develop an approach that captures both the diversity of thoughts and the extent to which an individual explores each thought category during pre-decision reflection.

\section{Framework}
In order to develop a framework to capture the diversity and depth of reflective thoughts, we adopted a theory-driven approach and grounded our work in previous reflection assessment models (Section \ref{sec:evalFrameworks}), specifically in Ullmann’s reflection model \cite{ullmann}. Grounded in a systematic analysis of 24 established reflection models, this model provides a robust foundation for our framework.
In Ullmann’s model, the categories are conceptualized as components of reflection, each serving as a quality marker. While this conceptualization is effective for assessing whether reflection reaches a quality threshold, it is less suited to our goal of examining reflective thought patterns in pre-decision contexts.
To align the framework with our purpose, we made three conceptual shifts:
\begin{itemize}
    \item \textbf{From components to aspects}. We reinterpreted the categories as aspects of reflective thinking rather than as components to be accumulated. This reinterpretation allowed each category to be treated as an independent aspect of reflection, making it possible to study variation in how people engage with different dimensions of reflection.
    
    \item \textbf{From binary coding to frequency}. Unlike Ullmann's framework which looks for the mere presence or absence of categories, we captured the frequency of occurrence of each aspect. This approach enabled us to measure the engagement with different reflective aspects, for example, whether an individual persistently revisits feelings, but rarely considers alternative perspectives.
    
    \item \textbf{From outcome-based depth to elaboration-based depth}. Ullmann defines depth in terms of whether reflection leads to an outcome, such as transformation or learning. We instead assessed the elaborateness of reasoning within each reflective aspect, drawing on concepts from deliberation studies \cite{Deliberation1}. This reconceptualization was crucial for pre-decision contexts, where reflection involves thinking and elaborating on thoughts rather than producing a final outcome.
\end{itemize}
The product of this step was the initial version of our framework (PROBE V0).
Then in a formative study we tested the applicability of this framework to pre-decision reflections, since the aspects in the framework were adapted from Ullmann's framework which has backward-looking orientation in contrast to pre-decision reflections with forward-looking orientation.
We first applied the framework to a small sample of pre-decision reflections, which highlighted the necessary revisions for this new context.
From this process, we introduced the final framework, PROBE V1, designed specifically to capture reflective thought patterns in pre-decision reflections. PROBE integrates various thought categories, their frequency of occurrence, and an elaboration-based depth measure into a single analytic lens.
In a summative study, we applied the final version of PROBE to a larger, more diverse dataset of pre-decision reflections, to demonstrate its reliability and its capability to uncover variation in reflective thought patterns across individuals.

\section{Method}
\begin{figure*}
\centering
\includegraphics[width=\textwidth]{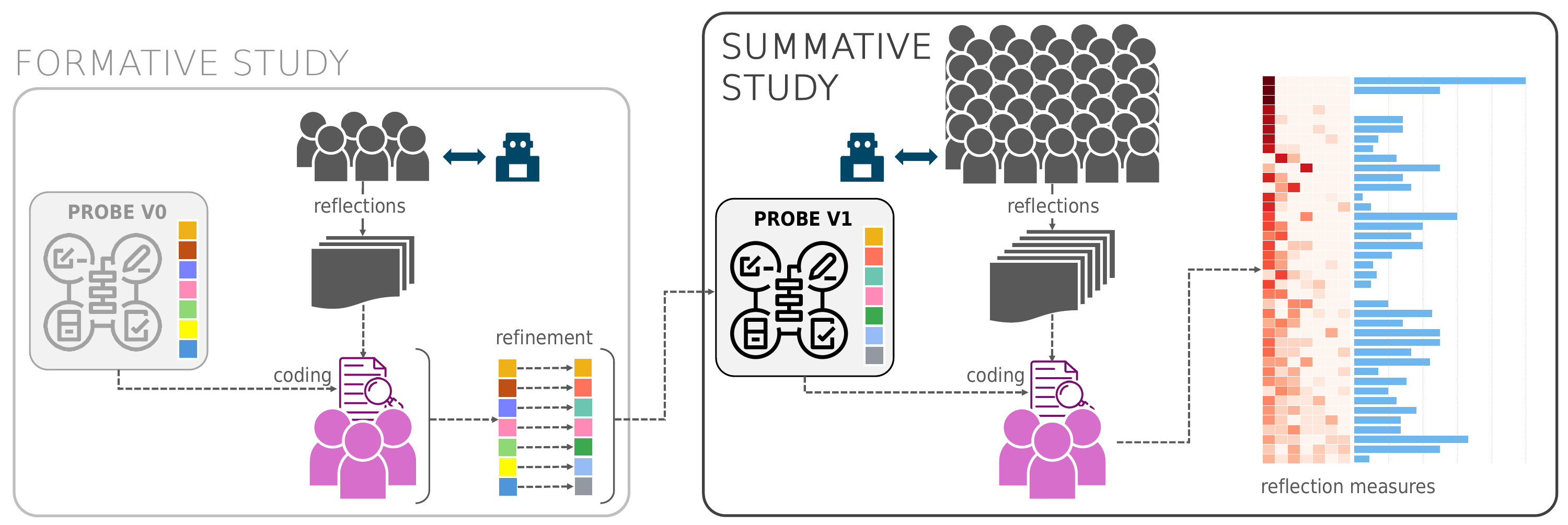}
\caption{Overview of the research study. The flowchart illustrates the progression from the Formative Study to the Summative Study. In the Formative Study, six participants’ pre-decision reflections (gathered through simple conversations with an agent) and the initial framework (PROBE V0) underwent iterative coding and refinement, during which PROBE categories were retained, slightly revised, or newly introduced. This process resulted in the refined framework (PROBE V1). The Summative Study then applied this framework to a larger set of participant reflections ($n=40$), which revealed diverse reflection patterns across participants. Together, the two phases highlight the process of framework development and evaluation, where empirical data informed refinement, and refined framework enabled systematic measurement.}
\label{fig:StudyOverview}
\Description{The figure is divided into two sections: the Formative Study (left) and the Summative Study (right).
Formative Study. The study begins with the initial framework, PROBE V0. Participants (group icon), together with an agent (robot icon), provide reflections, represented as data (stack of documents icon). These data, combined with the initial framework, move into a coding stage (three coders with magnifying glass icon). Coding then leads to a refinement stage (vertical column of colored boxes), where arrows indicate categories that were retained, slightly refined, or newly introduced. This stage produces a refined version of the initial PROBE framework.
Summative Study. An arrow connects the refinement stage to the Summative Study section. Here, additional participants provide reflections, which are analyzed through coding using the refined framework (PROBE V1).This process leads to reflection measures (illustrated as plots).}
\end{figure*}

This section reports on both, the formative and summative studies designed to help refine and evaluate PROBE respectively. 
The study design and data collection aspects were the same for both studies, which we report together in Section~\ref{sec:study_design}.
The subsequent sections detail the formative and then summative studies separately. The research process is also summarized in Figure \ref{fig:StudyOverview}.

\subsection{User study design \& data collection}
\label{sec:study_design}

We designed a user study in which participants generated written reflections on personally meaningful decisions. Writing was chosen as a method because it externalizes internal reflection while preserving participants’ own framing, allowing us to study PDR in a structured yet naturalistic way.

\textbf{Reflection Topics}.
To ensure individual relevance and diversity, reflection topics were selected from the category of big life decisions~\cite{camilleri2023investigation}, which are known to have significant long-term consequences and are moments when reflection is most needed~\cite{mols2016informing}. The topics were chosen to be sufficiently general for a broad demographic and align with those identified as important in prior work~\cite{camilleri2023investigation}. Table~\ref{table:topics} provides an overview of the decision topics and their categories.

\begin{table}
\centering
\begin{tabular}{ll} 
\toprule
\textbf{Decision Category} & \textbf{Decision Topic} \\ 
\midrule
Career & Choosing between a part-time or full-time job \\ 
Family & Choosing between having children or remaining child-free \\ 
Finances & Choosing between buying or renting a house \\ 
Relocation & Choosing between moving to a new city/country or staying in place \\ 
\bottomrule
\end{tabular}
\caption{Decision Topics and Categories}
\label{table:topics}
\end{table}

\textbf{Participants}.
We recruited 46 participants through the crowd-sourcing platform Prolific \footnote{\url{https://www.prolific.com}}, with ethics approval obtained from our university’s Human Research Ethics Committee (HREC).
Six of the 46 participants were recruited for the formative study, and the remaining 40 for the summative study.
To ensure diverse reflection styles, participants varied in age, gender, and educational background, and we assessed their self-reflection capability using the Self-Reflection and Insight Scale (SRIS)~\cite{short-SRIS}. This diversity allowed us to capture a broad range of perspectives relevant to our study.

\textbf{Procedure}. 
After providing informed consent detailing the study purpose, data use, and participant rights, individuals completed a pre-questionnaire.
The pre-questionnaire gathered demographics and included the Self-Reflection and Insight Scale (SRIS). Participants then selected one decision topic and reflected through a web-based conversational agent developed to elicit text-based reflections, with a future reflection support agent in mind.
To collect unmediated reflections, the agent remained passive, limited to introducing topics and encouraging detail.
An example interaction between the agent (which the users knew as ReflectiMate) and a participant is illustrated in
Figure \ref{fig:Interaction}.

After the reflection phase, participants completed a post-questionnaire on perceived thoroughness, for which we introduced two Likert-scale and open-ended questions in the absence of validated measures.
(1) "To what extent did you cover multiple factors of your personal context in your reflection?"
which assessed their subjectively perceived breadth of reflection
(2) "How deeply did you explore your personal context and thoughts around the choices?" 
which assessed perceived reflection depth, 
along with their reasoning captured in an open-ended response.

These measures were later analyzed in conjunction with coded reflections to triangulate insights.
Additionally, to inform future support design, the questionnaire sought to capture participants’ willingness to receive support from the agent in the reflection process and the reasons for that. 
The median time spent by participants on the study was 15 minutes and 4 seconds.

\textbf{Dataset}. In total, the study yielded 46 written reflections. A subset of six reflections was used in the formative study to refine the framework. A larger subset of data including 40 reflections subsequently analyzed in the summative study to evaluate PROBE’s reliability and variations in reflective patterns.

\subsection{Study 1: Formative study}
The formative study served as an exploratory validation step to examine the applicability of adapted framework (PROBE v.0) (see Table \ref{table:CS_initial}) to the domain of pre-decision reflection (PDR).
Three coders participated in the coding study, all of whom were early-stage researchers familiar with the concept of reflection and its incorporation into research.
As mentioned earlier, the coders coded reflections from the six participants recruited for this study.
The process followed by the coders is itemized below:

\textbf{Preparation:} Coders familiarized themselves with the framework and with the conceptual shifts we introduced (aspects instead of components, frequency instead of binary coding, elaboration-based depth instead of outcome-based depth) and were allowed to ask clarification questions from the researcher to fully understand it.

\textbf{Independent Application:} They independently coded the six reflections using the adapted categories in the Inception annotation platform \cite{inception}; and 

\textbf{Revision Identification:} They documented any ambiguities, mismatches, or limitations encountered.

\textbf{Consensus Building:} 
After independent coding, coders held a structured discussion to compare reflections, resolve disagreements, and refine the framework by addressing systematic challenges.

The above steps necessitated redefinition and refinement of some coding categories to align with the forward-looking nature of PDR, resulting in the final framework (see Table \ref{table:PROBE}).

\subsection{Study 2: Summative study}
The summative study served as a confirmatory evaluation of PROBE. It aimed to evaluate PROBE’s applicability and reliability on a larger and more diverse sample, and generate insights into reflective patterns across individuals with varying levels of self-reflectiveness.

\begin{figure}
\centering
\includegraphics[width=0.5\textwidth]{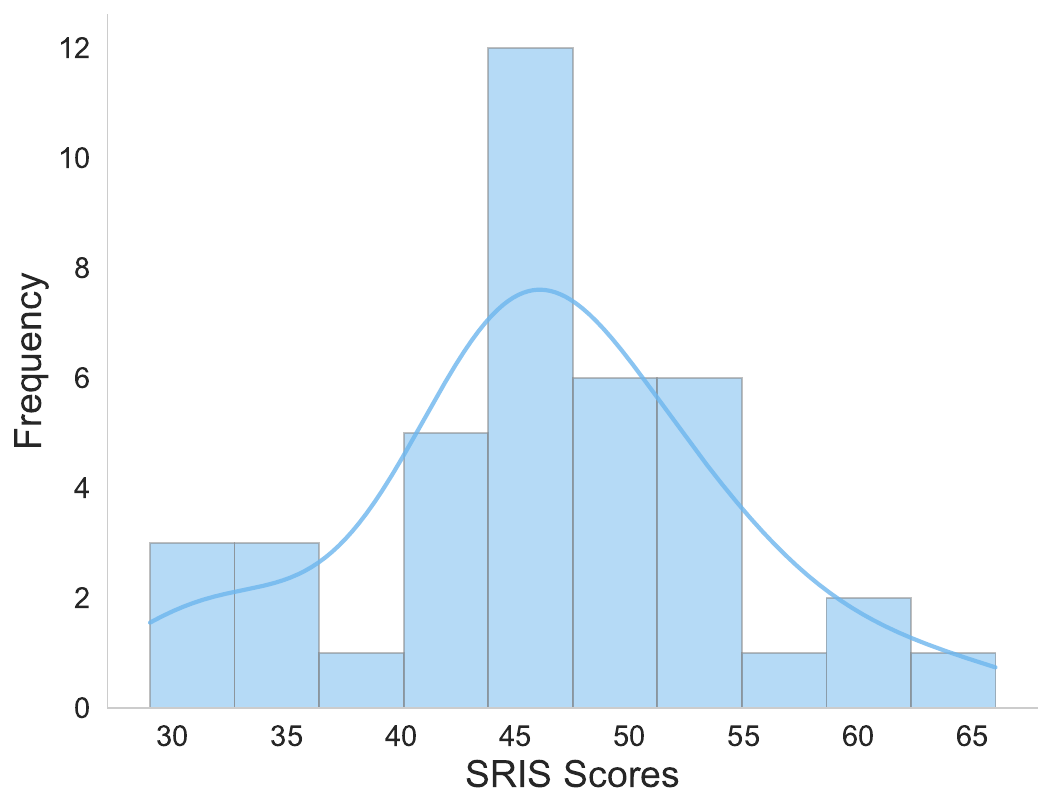}
\caption{Distribution of SRIS scores collected from participants ($n=40$) of the summative study. It is roughly bell-shaped, peaking between scores of 45 and 50. The mean score was 46.12 (SD = 8.28) on a scale with a maximum of 72.}
\label{fig:SRIS}
\Description{A histogram titled 'Distribution of SRIS Scores.' The x-axis is labeled 'SRIS Scores' and ranges from 30 to 65 in increments of 5. The y-axis is labeled 'Frequency' and ranges from 0 to 12 in increments of 2. There are eight vertical bars representing the frequency of scores within specific ranges.

Bar 1: From 30 to 35, the bar height is at 3.

Bar 2: From 35 to 40, the bar height is at 1.

Bar 3: From 40 to 45, the bar height is at 5.

Bar 4: From 45 to 50, the bar height is at 12, the tallest bar in the histogram.

Bar 5: From 50 to 55, the bar height is at 6.

Bar 6: From 55 to 60, the bar height is at 1.

Bar 7: From 60 to 65, the bar height is at 2.

Bar 8: From 65 to 70, the bar height is at 1.

A thick, smooth, blue curve is superimposed over the histogram bars. The curve rises from the low end of the scores, peaks between scores 45 and 50, and then falls towards the high end. This curve approximates a normal, or bell-shaped, distribution, showing that most scores are clustered around the center of the range, with fewer scores at the lower and higher ends.}
\end{figure}

\begin{figure}
\centering
\includegraphics[width=0.5\textwidth]{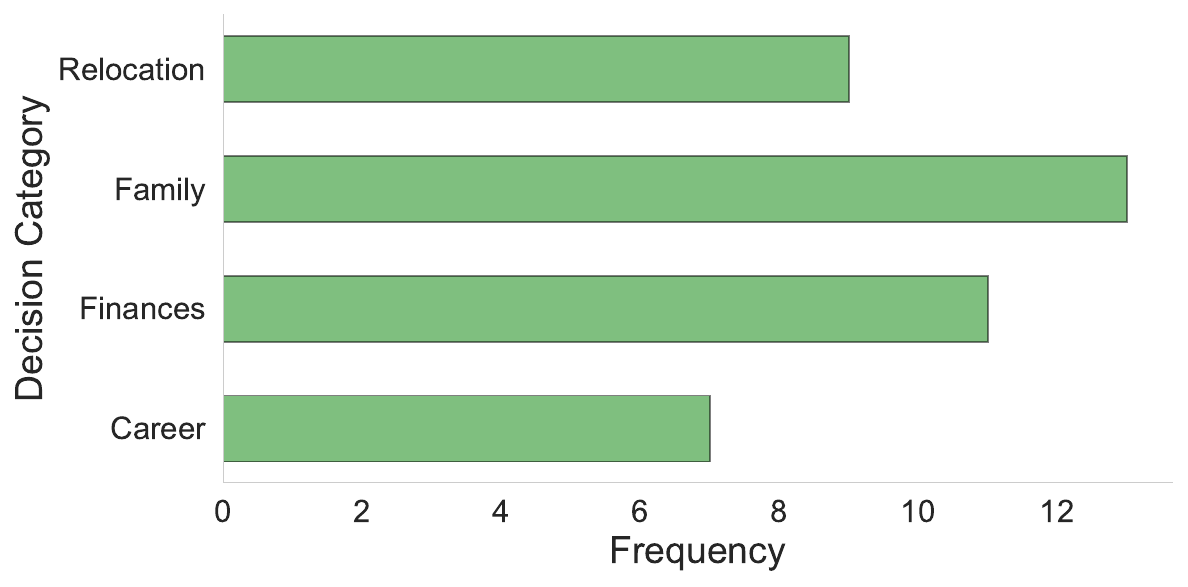}
\caption{Frequency of decision categories. The bar chart depicts the frequency of each decision category in the summative study data. \textit{"Career"} was chosen by 7 participants, \textit{"Finances"} by 11, \textit{"Family"} by 13, and \textit{"Relocation"} by 9.
The overall distribution suggests that the data is diverse in terms of decision topics.}
\label{fig:TopicFrequency}
\Description{Bar chart illustrating the frequencies of decision categories. The 'Family' category is the most frequent with 13 occurrences, followed by 'Finances' with 11, 'Relocation' with 9, and 'Career' with 7.}
\end{figure}

We analyzed pre-decision reflections collected from 40 adults (20 male, 20 female; M = 38.83 years, SD = 9.83; median = 39.5 years) whom reported varying levels of self-reflection tendency, measured beforehand in the pre-questionnaire using the Self-Reflection and Insight Scale (SRIS). As shown in Figure \ref{fig:SRIS}, SRIS scores followed a relatively normal distribution, ensuring coverage across the spectrum of reflectiveness, from individuals with low reflective tendencies to those who self-identify as highly reflective. The participants also varied in age, gender and educational background (see Table \ref{table:preQ}). The collected reflections likewise spanned a diverse set of personal decision topics as  illustrated in Figure \ref{fig:TopicFrequency}.

All reflections were analyzed using PROBE, by the same three coders from the formative study. The process involved:
(1) Coding all occurrences of each reflective categories 
(2) Assessing the elaborateness of each coded instance

To ensure coding consistency, the process was conducted blind. One-third of the reflections were triple-coded by all three coders, while the remaining two-thirds were double-coded by two coders. Inter-coder reliability was assessed using standard measures: Fleiss’ $\kappa$ = .69 (substantial agreement) for the triple-coded subset, and Cohen’s $\kappa$ = .79 (substantial agreement) for the double-coded subset.

\section{Results}
\subsection{Results of Formative Study}
In this section, we outline the results of our formative study, focusing on the revisions and refinements that led to the final version of the PROBE framework, which was subsequently used in the summative study.
The finalized PROBE framework comprises seven thought categories. Below, we describe the adjustments made to individual categories, along with the rationale for these changes.

\textbf{Feeling}. This category remained largely unchanged, except for a minor name revision from "Feelings" to "Feeling", reflecting our focus on capturing each emotional thought individually. The categories in PROBE are conceptualized as aspects rather than components, making the singular form more precise. 

\textbf{Alternative Perspective}. Originally labeled "Perspective", this category was renamed to "Alternative Perspective" to better reflect its intent: capturing thoughts that adopt perspectives beyond the self. In the context of pre-decision reflection, this involves widening the lens to consider how others might approach or evaluate the decision. This contrasts with retrospective reflection, where taking other perspectives typically refers to understanding how others involved in the same past experience might have perceived it.

\textbf{Belief}. This category captures segments of reflection where the individual expresses personal opinions, assumptions, or convictions about the decision at hand. The definition was clarified to emphasize its focus on first-person perspectives, and the label was shortened from "Personal Belief" to "Belief" for consistency with the revised "Alternative Perspective". Importantly, while retrospective reflections may also involve beliefs, in PROBE this category is not anchored to specific past experiences but instead reflects the person’s general views and assumptions related to the forthcoming decision.

\textbf{Awareness of Difficulties}. This category was refined to capture thoughts indicating awareness of potential obstacles or limitations that may influence the decision currently being considered. Its name was shortened from "Awareness of Difficulties-Critical Stance" to "Awareness of Difficulties" to remove unnecessary complexity. The distinction from retrospective reflection is notable: whereas retrospective models emphasize identifying problems in past experiences to improve future practice (a central feature of reflection-for-action), in pre-decision reflection the focus is on anticipating potential barriers before the decision is made.

\textbf{Intention}. This category captures reflective thoughts in which individuals articulate their intentions regarding the decision, including specific plans or intended actions. It was developed by revising the "Future Intention" category in the initial framework. The revision was necessary because intentions expressed in pre-decision reflections reflect both prospective outcomes of the reflection itself and pre-existing intentions mentioned during the process. To reflect this broader scope, the label was shortened from "Future Intention" to "Intention".

\textbf{Experience}. This is a newly introduced category that captures references to both direct personal experiences (past and present) and indirect or observed experiences (e.g., what participants have seen or heard from others’ lives). We chose the broader label "Experience" to reflect its scope. 
This differs from the initial "Description of an Experience" category, which implied recounting a single past personal experience as the subject of reflection. 

\textbf{Insight}. This new category emerged from the initial "Lessons Learned" category. We observed reflective thoughts in which individuals drew connections between their current situation and similar past experiences and attempted to integrate those insights into the ongoing decision making process. Such reflective thoughts observed were not aimed at formulating lessons from the past but at actively incorporating insights from past experiences into the ongoing process of making a decision. This integrative orientation distinguishes "Insight" from "Lessons Learned" and warrants its recognition as a new category in the framework.

The categories "Description of an Experience" and "Lessons Learned" were removed from the framework. We did not observe them in pre-decision reflections in our data, mainly because conceptually they align more with retrospective reflection. In pre-decision reflection, the experience itself is not the central subject of reflection, and the goal is not to learn lessons from the past but to contemplate on an upcoming choice. Table \ref{table:PROBE} provides a concise description of each PROBE category.

\subsection{Results of Summative Study}
\subsubsection{Reflective Thought Patterns}
Application of PROBE to 40 pre-decision reflections collected from diverse participants revealed distinct patterns of thought, varying levels of elaboration, and heterogeneous engagement across PROBE categories. 
\begin{figure*}
\centering
\includegraphics[width=0.9\textwidth]{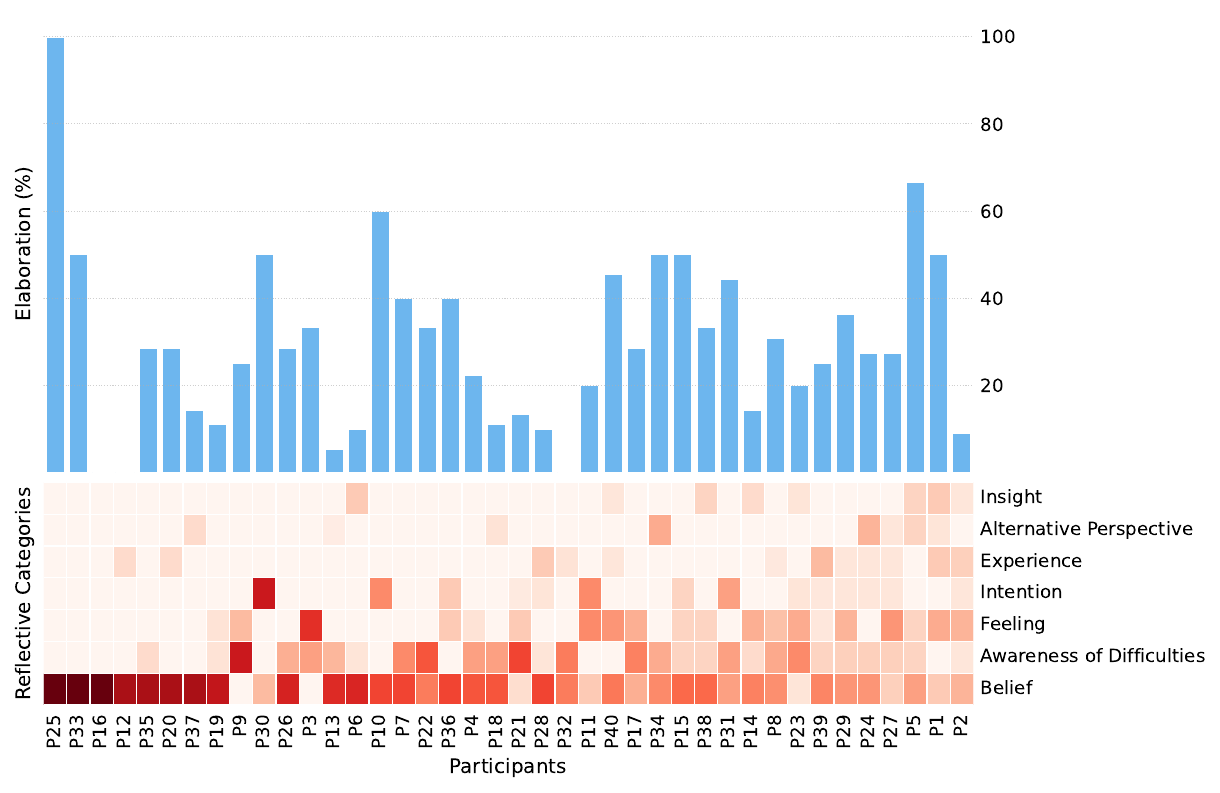}
\caption{Pre-decision Reflective Thought Patterns across participants from the summative study. The heatmap shows the distribution each participant's reflection across the seven PROBE categories, and the bar chart shows the depth of reflections for the same participants, defined as the percentage of elaborated thoughts among all their reflective thoughts. Columns represent participants, sorted by the variability of their reflections across categories (low variation on the left, high variation on the right), and rows are ordered by overall prevalence across all participants (most frequent at the bottom). The plots highlight the diversity of reflective patterns across participants, including distinctive cases such as P25, who exhibits very low diversity but high depth; P2, who shows high diversification but low elaboration; and P16, who shows not only very low diversification but also no elaboration.}
\label{fig:heatmap}
\Description{This is a combined data visualization containing a bar chart on the top and a heatmap on the bottom, both sharing a horizontal axis.

The horizontal axis is labeled "Participants," and it lists 40 unique participants, from P25 on the far left to P2 on the far right. The participants are not in numerical order.

The top part of the visualization is a 

bar chart with a vertical axis labeled "Elaboration (\%)" ranging from 0 to 100. Each participant has a single blue bar whose height indicates their elaboration percentage. The bars show a wide range of values, with the tallest bar belonging to P25, reaching nearly 100\%.

The bottom part is a  heatmap with a vertical axis labeled "Reflective Categories". The categories, from top to bottom, are Insight, Alternative Perspective, Experience, Intention, Feeling, Awareness of Difficulties, and Belief. The squares in the heatmap are shaded in different intensities of red, with darker shades indicating a higher presence of that reflective category for a given participant.

For example, participants P25, P33 have very dark red squares in the "Belief" row, indicating a high presence of this category}
\end{figure*}

Figure \ref{fig:heatmap} provides an overview of category-level engagement and elaboration levels per participant. 
As shown, participants differed considerably in the breadth of categories they engaged with. Some participants (e.g., P25, P33, P16) confined their reflections to a single category ("Belief") representing minimal diversification. 
Other participants engaged with a broader range of categories, though the extent of diversification varied considerably. Across all participants, the number of categories represented in reflections ranged from one to six (M = 3.2, SD = 1.34), although none engaged with all seven categories. 
This indicates a medium overall level of diversification.
Importantly, treating each PROBE category as an aspect of reflective thinking rather than a component enabled meaningful identification of these patterns.

Engagement was also uneven across categories both within and between participants. Some participants (e.g., P1, P2) distributed their attention relatively evenly across multiple categories, while others (e.g., P28, P30, P21) despite diversification, showed strong concentration in a single category, such as ``Awareness of Difficulties'' in P21. These individual differences reveal the distinct patterns through which participants approached their decisions, highlighting their most focused areas and areas where targeted support may be necessary to encourage diversification. Such insights were enabled by analyzing the relative frequency of categories rather than their mere presence or absence in our approach.

Despite these differences, consistent tendencies emerged across the dataset. The most prevalent category was ``Belief'', indicating that nearly all participants grounded their reflections in personal convictions or assumptions. ``Awareness of Difficulties'' was the second most frequent, indicating frequent recognition of challenges, followed by ``Feeling'', ``Intention'', and ``Experience''.
By contrast, ``Alternative Perspective'' and ``Insight'' were least represented, suggesting that these categories may be particularly difficult for participants to engage with and thus warrant targeted support.
The heatmap in Figure~\ref{fig:heatmap} shows the distribution of reflection categories across participants.

The bar chart in Figure~\ref{fig:heatmap} shows the variation in the elaboration of thoughts across participants.
While P25 provided elaborations for all reflective thoughts, others (e.g., P16, P12, P32) produced exclusively non-elaborated reflections. 
Table~\ref{table:thoughts} provides examples of elaborated and non-elaborated thoughts.
Most of the participants ($\sim$80\%) elaborated fewer than half of their thoughts in their reflections.
This suggests an overall low level of depth in pre-decision reflection and a need for elaboration support.
The elaboration-based depth measure in PROBE can identify which specific thoughts lack elaboration, allowing for targeted depth-enhancing interventions.

Finally, combining insights on breadth (diversity) and depth revealed distinct participant profiles. For example, P25 demonstrated high depth (100\% elaboration) but no diversity in reflection aspects, whereas P2 showed high diversity but very limited elaboration. Conversely, P16 exhibited both low diversity and low depth.
Such patterns highlight how PROBE provides a structured overview of individual reflective processes, enabling identification of where support is most needed, whether to foster more diversification, deeper elaboration, or both.

\subsubsection{Self-reported Assessment}
\begin{figure*}[t]
\centering
\includegraphics[width=\textwidth]{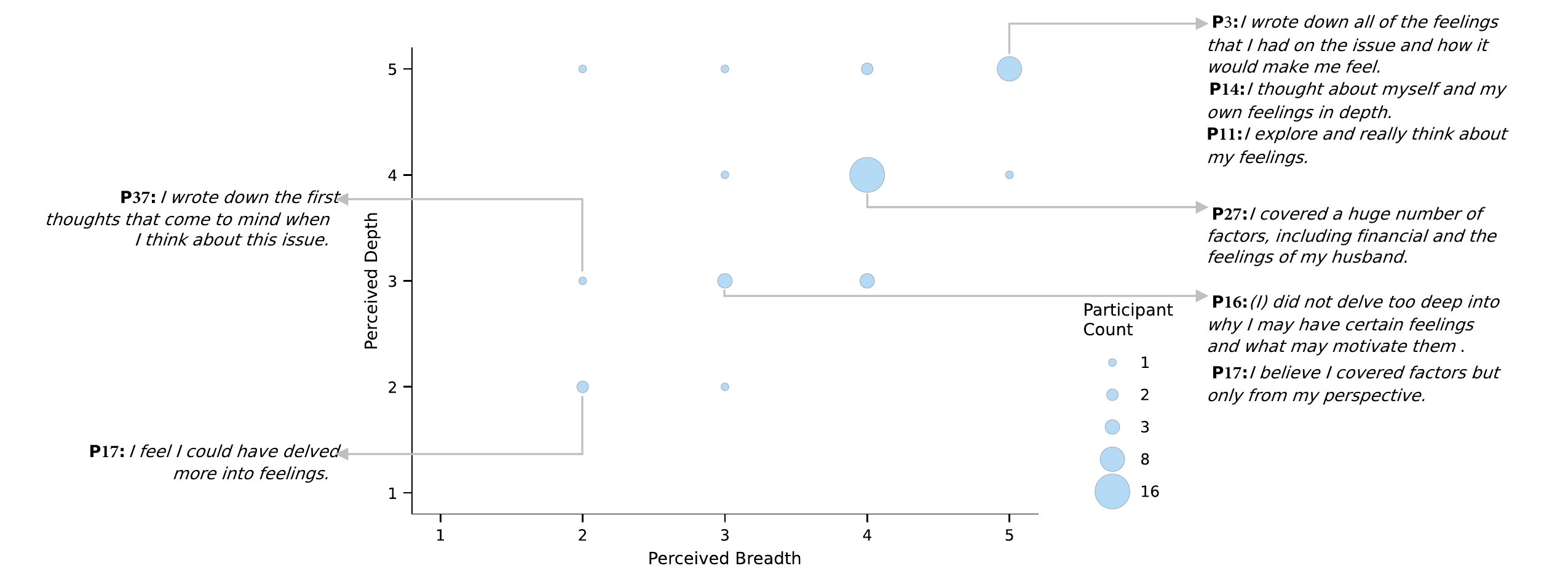}
\caption{Distribution of participants by their self-assessments of the breadth and depth of their own reflections on a five-point Likert scale. The size of the circles indicate the number of participants in the corresponding depth vs.\ breadth ``bin''. Quotes from participants' self-assessments of their reflection show a focus on the process of reflection, but lack specifics on the quality and aspects of reflection.}
\label{fig:Comments}
\Description{The figure is a bubble scatter plot with "Perceived Depth" on the y-axis and "Perceived Breadth" on the x-axis, both ranging from 1–5. Bubble size represents participant count: small = 1, medium = 2–3, large = 8, largest = 16.
16 participants: Breadth 4, Depth 4
8 participants: Breadth 5, Depth 5
3 participants: Breadth 3, Depth 3; Breadth 4, Depth 3
2 participants: Breadth 4, Depth 5; Breadth 2, Depth 2
1 participant each: Breadth/Depth pairs (2,3), (2,5), (3,2), (3,4), (3,5), (5,4)
Participant comments are placed around the graph:
High breadth & depth (5,5): P3, P14, P11 describe deeply reflecting on their feelings.
Breadth 4, Depth 4: P27 notes considering many factors, including financial and relational.
Breadth 3, Depth 3: P16 and P17 mention limited exploration, mostly from their own perspective.
Breadth 2, Depth 3: P37 notes writing initial thoughts only.
Breadth 2, Depth 2: P17 again notes they could have gone deeper into feelings.}
\end{figure*}

Despite the varied thought patterns captured by PROBE, participants’ self-rated breadth and depth of reflection, collected via a post-questionnaire on a 5-point Likert scale, showed very little variation (Figure \ref{fig:Comments}). Ratings of breadth and depth tended to increase in tandem and were often identical, suggesting that participants perceived the two dimensions as closely related. Only one participant diverged from this trend, reporting low breadth (2) but maximum depth (5). Overall, participants rated both dimensions relatively highly, with the most frequent response being 4 for both breadth and depth (16 participants). The mean and standard deviation values (Breadth: M = 3.875, SD = 0.88; Depth: M = 3.975, SD = 0.89) indicate that participants generally had confidence in the comprehensiveness and thoroughness of their pre-decision reflections.

Open-ended explanations collected in the post-questionnaire provided further context for participants’ self-assessments. Several participants (e.g. P3, P14, P11) who rated both breadth and depth at the maximum level (Breadth = 5, Depth = 5) emphasized reflecting on their feelings (see Figure~\ref{fig:Comments}). In addition, some participants (e.g. P16 and P17) who rated their depth relatively lower attributed this to limited engagement with their feelings. This suggests that, for several participants, perceived breadth and particularly perceived depth, was closely tied to the extent of emotional reflection.

High ratings of breadth and depth were also associated with the inclusion of personally salient factors. For instance, P27 justified their high self-ratings by highlighting the consideration of two key factors that were especially meaningful to them (financial aspects and their husband’s feelings).

Another observation is the general lack of acknowledgment regarding the possible limitations or narrowness of the reflections. Only a few participants explicitly mentioned some limitations. For example, P37 noted that their reflections were based only on immediate accessible thoughts. P17 acknowledged that they were confined to a first-person perspective and could have included alternative perspectives (see Figure \ref{fig:Comments}). 
However, most open-ended responses were general and non-specific (e.g., ``it was deep enough,'' ``I reflected as much as I can''), limiting the extent to which these explanations provide insights beyond participants' confidence in their cognitive processes at this stage.

Despite the lack of explicit acknowledgment of limitations in their justifications and the generally high scores they reported, the majority of participants (80\%) indicated in their post-reflection responses a preference for support from a conversational agent, primarily to help identify overlooked thoughts or encourage deeper consideration during reflection.
Below are some illustrative excerpts that reflect the motivations.

\begin{quote}
    ``It can be helpful to have guidance in reflection—the way a therapist would ask questions to help navigate the various
    issues.'' \textit{(P37)}
\end{quote}
\begin{quote}
    ``It could have helped focus on a specific aspect I maybe brushed over and could have thought more about.'' \textit{(P27)}
\end{quote}
\begin{quote}
    ``It would be good to see if there are things that I haven’t thought about in that scenario, as it may change my perspective.''
\textit{(P28)}.
\end{quote}
\begin{quote}
    ``I think guided questioning would encourage deeper thinking.'' \textit{(P16)}
\end{quote}

\subsubsection{Reflective Thoughts}
To illustrate the characteristics of pre-decision reflective thoughts, Table~\ref{table:thoughts} presents some examples from multiple participants and topics, organized by PROBE categories and divided into elaborated or not elaborated. As shown, elaborated thoughts incorporate reasoning and justification, whereas non-elaborated thoughts remain surface-level.

\begin{table}[th!]
\small
\newcolumntype{L}[1]{>{\raggedright\let\newline\\\arraybackslash\hspace{0pt}}p{#1}}
\centering
\begin{tabular}{L{1.8cm} L{5.8cm} L{6.4cm}} 
\toprule
\textbf{Category}  & \textbf{Not Elaborated} & \textbf{Elaborated} \\ 
\midrule
\textbf{Belief} & Working full-time provides a comfortable lifestyle and gives a sense, rightly or wrongly, of security. \newline \textit{(P6), Topic: Choosing between a part-time job or a full-time job} 
& Ever since I was a child myself, I never really wanted to start a family of my own because always wanted to go traveling and focus on my career. \newline
\textit{(P2), Topic: Choosing between having children or remaining child-free} \\[0.75em]
\cmidrule(lr){1-3}
\textbf{Awareness of Difficulties } & The housing market is so expensive right now.  \newline \textit{(P21), Topic: Choosing between buying or renting a house} 
& Renting a house is challenging for us because we have pets, and not everyone allows them in rented properties, whereas buying, you don’t have any issue. \newline \textit{(P21), Topic: Choosing between buying or renting a house} \\[0.75em]
\cmidrule(lr){1-3}
\textbf{Insight} & The pressure to conform and climb the greasy pole has had a serious impact on my physical and mental health.  \newline \textit{(P6), Topic: Choosing between a part-time job or a full-time job} 
& It took us almost 18 months to go through the selling and buying process and caused so much stress and heartache that I swore I would never do it again.  \newline \textit{(P23), Topic: Choosing between moving to a new city/country or staying in place} \\[0.75em]
\cmidrule(lr){1-3}
\textbf{Alternative Perspective} & My husband, however is very keen to have children  \newline \textit{(P27), Topic: Choosing between having children or remaining child-free} 
& Some people prefer renting over buying to have the freedom to move, not only to a different local destination but if they want to travel abroad whether that be through work commitments or personal holiday then renting allows that freedom without being tied down.   \newline \textit{(P5), Topic: Choosing between buying or renting a house} \\[0.75em]
\cmidrule(lr){1-3}
\textbf{Experience} & I have a brother and sister who are very happy and have made new lives in Australia and Canada.  \newline \textit{(P29), Topic: Choosing between moving to a new city/country or staying in place} 
& We have nieces and nephews that we look after frequently and this is enough for us, my husband has a terminal illness and we would not have the time to dedicate to a child.  \newline \textit{(P28), Topic: Choosing between having children or remaining child-free} \\[0.75em]
\cmidrule(lr){1-3}
\textbf{Intention} & I am thinking about leaving the UK and moving back to my hometown in Italy.  \newline \textit{(P15), Topic: Choosing between moving to a new city/country or staying in place} 
& I need to start looking around to see where is the best savings account with the best interest rate so that I can build up my savings quickly.  \newline \textit{(P30), Topic: Choosing between buying or renting a house} \\[0.75em]
\cmidrule(lr){1-3}
\textbf{Feeling} & The idea of living in the same place forever is really depressing to think about for me.  \newline \textit{(P40), Topic: Choosing between moving to a new city/country or staying in place} 
& I’m also a very stressed out person, I panic about everything, so the idea of having a child that is solely dependent on me scares me.  \newline \textit{(P8), Topic: Choosing between having children or remaining child-free} \\[0.75em]
\bottomrule
\end{tabular}
\caption{Pre-decision Reflective Thought Examples (From Participants of Summative Study)}
\label{table:thoughts}
\end{table}

\section{Discussion}
In this paper, we set out to devise a framework that renders the thought patterns behind life changing decisions both visible to users and computable by a system. Rather than prescribing what counts as a good or bad choice, our goal was to foster metacognitive awareness, supporting decision makers in recognizing which patterns shaped their reasoning and which thought patterns they may have left aside.

To this end we devised PROBE (Pre-decision Reflections fOr Big Life dEcisions). PROBE is the first framework to conceptualize decision making as forward-looking reflection. 
Our coding study demonstrated that coders reached high inter-rater reliability, showing that reflection-based approaches can be systematically applied to decision making.

We argue that this empirically tested theoretical conceptualization opens the door to a new class of decision-support tools. While prior systems have often emphasized contrasting perspectives or counterarguments (e.g. in \cite{DSS1,DSS2}), our approach extends this space by fostering broader metacognitive awareness, supporting decision makers in understanding the patterns that shape their reasoning beyond binary framings.

From the perspective of reflection support systems, this reframing also opens the door for new kinds of support tools. A well-known caveat of post-hoc reflection is the tendency toward rationalization, where people construct explanations to justify actions they have already taken~\cite{festinger1957cognitive, nisbett1977telling}. 
We define the term ``Pre-Decision Reflection'' (PDR) to anchor our approach toward reflection assessment, so that decisions may be \textit{informed} by---rather than interpreted after the fact by---the decision-maker's reasoning.

\subsection{Insights from Summative Evaluation}
In order to evaluate the framework further empirically, our evaluation of PROBE thus focused on (a) its ability to highlight the aspects or dimensions of pre-decision reflection as an indicator of the ``breadth'' of reasoning by the decision-maker, (b) its ability to highlight the degree of engagement and elaboration as an indicator of the ``depth'' of reasoning within individual aspects

With regards to (a) and (b) our study reveals that PDR was highly heterogeneous across participants:
Some participants reflected broadly across multiple aspects but failed to elaborate their thoughts; some participants reflected deeply but narrowly, elaborating their thoughts in one or two reflection aspects but failing to consider others; while a minority of participants achieved a reasonable balance across breadth and depth of reflections.
While some categories were readily engaged (e.g., ``Belief''), others such as ``Insight'' or ``Alternative perspective'' seem likely to benefit more from external support. Overall, depth was limited, 80\% of reflections included elaboration in fewer than half of the thoughts, highlighting clear opportunities for systems to scaffold deeper and broader engagement.

Post-reflection feedback echoed these findings: A few participants noted their reflections were shaped by immediate, personally salient thoughts, overlooking other considerations. This aligns with bounded awareness~\cite{bounded, bounded1}, where attention narrows to the familiar or salient. The shallowness of reasoning also resonates with Kahneman’s dual-process theory~\cite{fastslow}, suggesting that reflections were guided by fast, intuitive (System 1) thinking rather than slower, more effortful reasoning (System 2).

\begin{figure*}[t] 
  \centering
  
  \begin{subfigure}[t]{0.48\textwidth}
    \centering
    \includegraphics[width=\linewidth]{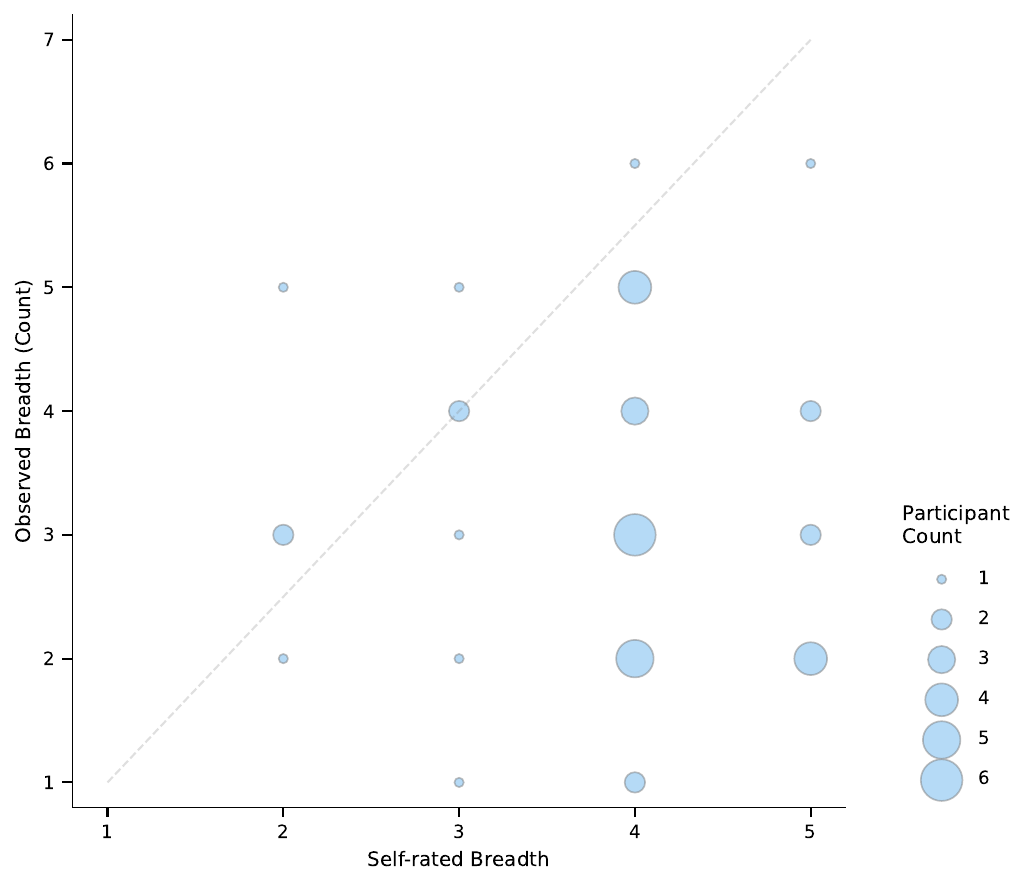}
    \caption{Observed vs Self-rated Breadth}
    \label{fig:breadth}
  \end{subfigure}
  \hfill
  \begin{subfigure}[t]{0.48\textwidth}
    \centering
    \includegraphics[width=\linewidth]{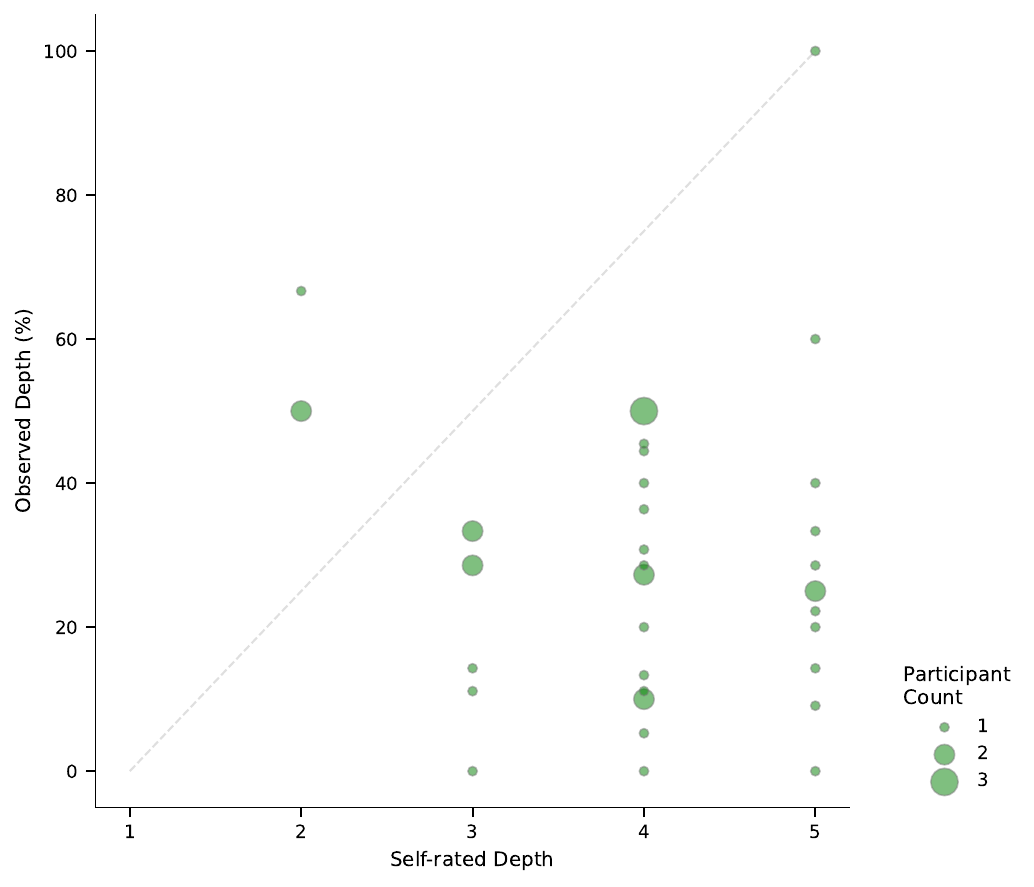}
    \caption{Observed vs Self-rated Depth}
    \label{fig:depth}
  \end{subfigure}
  
  \caption{Comparison of self-rated and observed measures. (a) Breadth, (b) Depth. Two bubble charts show the relationship between participants’ self-reported measures (collected post-reflection) and the corresponding breadth and depth of reflection captured with PROBE. Observed breadth is the number of PROBE categories present in a participant’s reflection, and observed depth is the percentage of elaborated thoughts. The dashed diagonal line in each chart represents perfect agreement between observed and self-reported values. The points above the diagonal indicate underestimation, while those below indicate overestimation.
  An overall tendency toward overestimation is evident in both plots, but it is more pronounced in depth.}
  \Description{This visualization displays two separate bubble scatter plots side-by-side, labeled (a) and (b).

Plot (a) is titled "Observed vs Self-rated Breadth." The horizontal axis is "Self-rated Breadth" on a scale from 1 to 5. The vertical axis is "Observed Breadth (Count)" on a scale from 0 to 7. A dashed diagonal line extends from the bottom left to the top right of the plot, representing where self-rated and observed breadth are equal. Several light blue bubbles are plotted, with the size of each bubble corresponding to the number of participants at that specific data point, as shown in a small legend. For example, a large bubble at x-axis 4 and y-axis 3 indicates that six participants had a self-rated breadth of 4 and an observed breadth of 3.

Plot (b) is titled "Observed vs Self-rated Depth." The horizontal axis is "Self-rated Depth" on a scale from 1 to 5. The vertical axis is "Observed Depth (\%)" on a scale from 0 to 100. A dashed diagonal line also extends from the bottom left to the top right, representing where self-rated and observed depth are equal. Several green bubbles are plotted, with the size of each bubble corresponding to the number of participants at that data point. The bubbles in this plot show a noticeable cluster below the diagonal line.}
  \label{fig:selfobserved}
\end{figure*}

Interestingly, participants rated the depth and breadth of their unassisted reflections higher than PROBE measures would indicate (see Figure~\ref{fig:selfobserved}). This discrepancy may stem from overconfidence in self-assessment, as noted by Efklides~\cite{efklides, efklides2}, or from crowd workers rating themselves favorably to present their work positively. These findings underscore the need to triangulate self-reports with independent measures in future system design and research.

\subsection{Implications for Decision Support Systems}
While PROBE could be embedded within black-box systems, we designed it to serve as an important component in future system designs that can facilitate users’ agency over their own decision-making.
We envision systems that not only support users longitudinally across multiple decision scenarios but also help them remember and meta-reflect across those experiences. This reframing also opens the door to dialogue: inviting users to articulate how they would like to be supported by AI and enabling co-creation around both observed and perceived patterns.
Below we provide some examples.

From a system designer perspective this has many implications.
PROBE could be used in many different manners.
First, designers of decision support systems can use PROBE to provide users with a mirror of their own thought patterns. Our results show that people are often unaware of the limits of their reflections or tend to judge the quality of their reflections based on emotional engagement rather than an awareness of breadth. Raising this awareness—echoing findings from behavioral support systems, can be a crucial first step toward change.

PROBE was also designed with future adaptation for agentic AI in mind. A conversational system could prompt users to explore different kinds of thought patterns, for example by noting 
``your reflections seem to be based on your convictions; perhaps thinking of all the difficulties you anticipate might give you a different perspective?''
or by actively posing questions that target specific categories. 
Applying PROBE in agentic AI settings will shift these systems from steering users towards specific outcomes to leveraging their rational strengths to support the process of pre-decision reflection. This positions AI in a safer space, where clear metrics guide exploration of diverse thought patterns rather than premature solutions.

We found strong adoption potential: most participants welcomed assistance during pre-decision reflection, unlike earlier work where many preferred to reflect alone~\cite{mols2016informing}. This suggests a particular need for decision-support in big life choices, where the risks of shallow reflection are higher, and may also reflect shifting attitudes toward AI, participants’ growing familiarity with LLM-powered conversational agents 
appears to increase willingness to accept active, prompting support. Together, these trends open a clear opportunity for designers to create tools that aid people in navigating high-stakes decisions in sensitive but effective ways.

With regard to the use of PROBE to assess the quality of pre-decision reflections, our stance is to avoid making judgments about the quality of decision making processes. While certain domains may allow clearer right–wrong distinctions, big life decisions are deeply individual, shaped by personal circumstances that influence pre-decision thoughts. Our aim is not to prescribe but to ensure that our thought categories sufficiently capture the reflections observed in our study. Although the sample size could always be expanded, the framework itself is inherently extendable and designed to accommodate broader application.

\subsection{Limitations and Future Work}
Our approach assumes that breadth categories can be reliably detected—ideally automatically—though this may not always hold, even if advances in NLP and large language models make it increasingly feasible. Treating all categories equally and relying on a limited, demographically constrained sample may also restrict generalizability, highlighting the need for larger and more diverse studies.
We also recommend that future research explore how the relative importance of different categories can be integrated into the proposed framework, guided by theoretical foundations or expert input.
The automatic identification of PROBE categories should also be explored as an initial step toward real-time deployment in support systems.

\section{Conclusion}
In this paper, we conceptualized decision making as a forward-looking reflective activity and introduced PROBE, the first framework designed to capture the thought patterns shaping pre-decision reflections. PROBE operationalizes reflection along two dimensions: (1) breadth, indicating the scope of reflective thoughts, and (2) depth, indicating their elaborateness. 
Applying PROBE allowed us to quantify pre-decision reflections.
Overall, our findings show that PROBE is a reliable and versatile tool for uncovering thought patterns. The identified patterns reveal substantial heterogeneity across participants, highlighting that many individuals could benefit from support in both diversifying and deepening their thinking.

Notably, participants perceived their unassisted reflections as deeper and broader than PROBE’s quantification suggested, underscoring the value of combining subjective experience with systematic assessment. Taken together, our findings highlight PROBE’s potential for systems that not only foster self-awareness but also strengthen people’s agency in choosing which thought patterns to rely on for their high-stakes decisions. By making hidden thought patterns visible, PROBE opens new opportunities for technologies that actively expand how people reflect, reason, and decide.


\begin{acks}
We thank Mohammed Al Owayyed and Michaël Grauwde for their contribution to the construction of the PROBE. 
This research was (partially) funded by the Hybrid Intelligence Center, a 10-year programme funded by the Dutch Ministry of Education, Culture and Science through the Netherlands Organisation for Scientific Research, https://hybridintelligence-centre.nl, grant number 024.004.022.
In addition, it has been partially supported by the BOLD cities initiative.
\end{acks}

\bibliographystyle{ACM-Reference-Format}
\bibliography{main}

\newpage
\appendix
\section{Appendix}
\label{sec:appendix}

\subsection{Reflection Aspects in V0 \& V1 versions of PROBE}

\begin{table}[h!]
\renewcommand*{\arraystretch}{1.5}
\newcolumntype{L}[1]{>{\raggedright\let\newline\\\arraybackslash\hspace{0pt}}p{#1}}
\centering
\begin{tabular}{L{4.5cm}L{8cm}} 
\toprule
\textbf{Category}  & \textbf{Indicator} \\ 
\midrule
\textbf{Personal belief} & The reflector describes his or her beliefs \\
\textbf{Awareness of difficulties-\newline Critical Stance} & The reflector recognizes and describes difficulties/problems \\
\textbf{Description of an experience} & The reflector describes an experience she or he had in the past \\
\textbf{Feelings} & The reflector describes his or her feelings and emotions \\
\textbf{Future Intentions} & The reflector intends to do something\newline (Prospective outcomes of the reflection process) \\
\textbf{Lessons Learned} & The reflector has learned something\newline (Retrospective outcomes of the reflection process)  \\
\textbf{Perspective} & The reflector takes into account another perspective\newline (from the person's own) \\
\bottomrule

\end{tabular}

\caption{Initial Framework (PROBE V0)}
\label{table:CS_initial}
\end{table}

\begin{table}[h!]
\renewcommand*{\arraystretch}{1.5}
\newcolumntype{L}[1]{>{\raggedright\let\newline\\\arraybackslash\hspace{0pt}}p{#1}}
\centering
\begin{tabular}{L{4.5cm}L{8cm}} 
\toprule
\textbf{Category}  & \textbf{Indicator} \\ 
\midrule
\textbf{Belief} & Captures personal opinions, assumptions, or convictions \\
\textbf{Awareness of difficulties} & Captures thoughts acknowledging (potential) obstacles\newline or limitations \\
\textbf{Experience} & Captures thoughts referring to direct or observed\newline personal experiences \\
\textbf{Feeling} & Captures individual emotional thoughts \\
\textbf{Intention} & Captures thoughts about plans or intended actions,\newline including both pre-existing and emerging intentions \\
\textbf{Insight} & Captures thoughts that integrate past experiences into the ongoing decision making process \\
\textbf{Alternative perspective} & Captures thoughts adopting perspectives beyond the self \\
\bottomrule
\end{tabular}

\caption{Final PROBE Framework (PROBE V1)}
\label{table:PROBE}
\end{table}

\newpage
\subsection{Summative Study: Additional Details}

\begin{table}[h]
\centering
\begin{tabular}{l l r}
\toprule
\textbf{Demographic} & \multirow{2}{*}{\textbf{Category}} & \textbf{Participant} \\
\textbf{Variable} &  & \textbf{Count}  \\
\midrule
\textbf{Age Range} & 18--24 & 2 \\
 & 25--34 & 12 \\
 & 35--44 & 16 \\
 & 45--54 & 7 \\
 & 55--64 & 3 \\
\cmidrule(lr){1-3}
\textbf{Education} & Doctorate degree or higher & 2 \\
 & Master's degree & 7 \\
 & Bachelor's degree & 18 \\
 & Associate degree & 1 \\
 & Some college credit, no degree & 5 \\
 & High school graduate & 7 \\
\bottomrule

\end{tabular}
\caption{Participant demographics for the summative study (collected through pre-questionnaire).}
\label{table:preQ}
\end{table}

\begin{figure*}[h]
\centering
\includegraphics[width=0.8\textwidth]{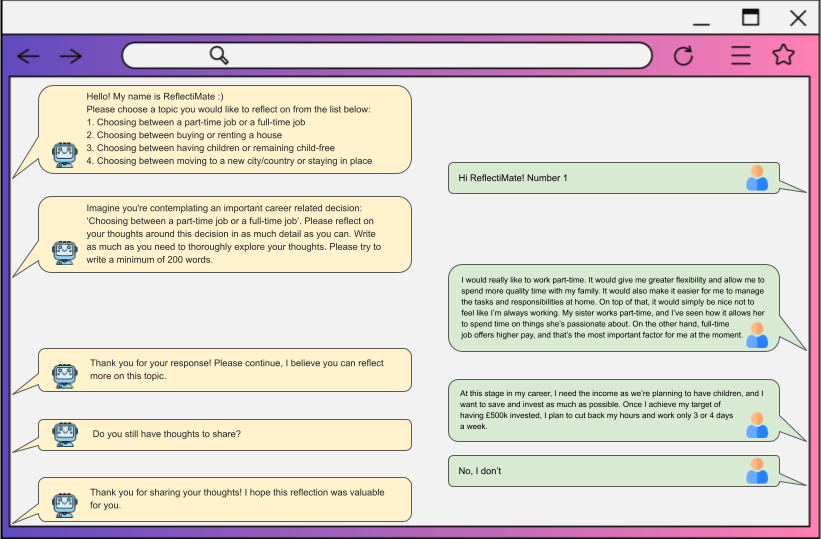}
\caption{Interaction Environment and Example Between Conversational Agent and Participant}
\label{fig:Interaction}
\Description{This figure illustrates a dialogue interface between a participant and the agent named "ReflectiMate." The visual includes the following:
Interface Design:
A browser-like window
Initial Prompt:
ReflectiMate provides a list of four topics for reflection:
Choosing between a part-time job or a full-time job.
Choosing between buying or renting a house.
Choosing between having children or remaining child-free.
Choosing between moving to a new city/country or staying in place.
The user selects the first option: “Choosing between a part-time job or a full-time job.”
Next Agent Prompt:
Imagine you're contemplating an important career related decision: ‘Choosing between a part-time job or a full-time job’. Please reflect on your thoughts around this decision in as much detail as you can. Write as much as you need to thoroughly explore your thoughts. Please try to write a minimum of 200 words.
User Responses:
I would really like to work part-time. It would give me greater flexibility and allow me to spend more quality time with my family. It would also make it easier for me to manage the tasks and responsibilities at home. On top of that, it would simply be nice not to feel like I’m always working. My sister works part-time, and I’ve seen how it allows her to spend time on things she’s passionate about. On the other hand, full-time job offers higher pay, and that’s the most important factor for me at the moment.
Agent Prompt:
Thank you for your response! Please continue, I believe you can reflect more on this topic.
User Responses:
At this stage in my career, I need the income as we’re planning to have children, and I want to save and invest as much as possible. Once I achieve my target of having £500k invested, I plan to cut back my hours and work only 3 or 4 days a week.
Agent Prompt:
Do you still have thoughts to share?
User Responds:
No, I don’t
Agent Answers:
Thank you for sharing your thoughts! I hope this reflection was valuable for you.}
\end{figure*}

\end{document}